\documentclass[reprint,aip,jcp,nobalancelastpage]{revtex4-2}

\usepackage{amsfonts}
\usepackage{amsmath}
\usepackage{mathtools}
\usepackage{siunitx}
\usepackage{graphicx}
\usepackage{bbm}
\usepackage{float}
\usepackage{algorithm}
\usepackage{algorithmicx}
\usepackage{algpseudocode}

\DeclareMathOperator{\E}{\mathbbm{E}}

\DeclarePairedDelimiter{\set}{\{}{\}}

\DeclarePairedDelimiterX{\inner}[2]{\langle}{\rangle}{{#1},{#2}}

\usepackage{lineno,hyperref}
\modulolinenumbers[1]

\begin{document}
\title{BAD-NEUS: Rapidly converging trajectory stratification}
\author{John Strahan}
\thanks{These authors contributed equally.}
\affiliation{Department of Chemistry and James Franck Institute, University of Chicago, Chicago IL 60637, USA}
\author{Chatipat Lorpaiboon}
\thanks{These authors contributed equally.}
\affiliation{Department of Chemistry and James Franck Institute, University of Chicago, Chicago IL 60637, USA}
\author{Jonathan Weare}
\affiliation{Courant Institute of Mathematical Sciences, New York University, New York, New York 10012, USA}
\author{Aaron R. Dinner}
\email{dinner@uchicago.edu}
\affiliation{Department of Chemistry and James Franck Institute, University of Chicago, Chicago IL 60637, USA}
\begin{abstract}
An issue for molecular dynamics simulations is that events of interest often involve timescales that are much longer than the simulation time step, which is set by the fastest timescales of the model.  Because of this timescale separation, direct simulation of many events is prohibitively computationally costly. This issue can be overcome by aggregating information from many relatively short simulations that sample segments of trajectories involving events of interest. This is the strategy of Markov state models (MSMs) and related approaches, but such methods suffer from approximation error because the variables defining the states generally do not capture the dynamics fully.  By contrast, once converged, the weighted ensemble (WE) method aggregates information from trajectory segments so as to yield unbiased estimates of both thermodynamic and kinetic statistics.  Unfortunately, errors decay no faster than unbiased simulation in WE as originally formulated and commonly deployed.  Here we introduce a theoretical framework for describing WE that shows that introduction of an approximate stationary distribution on top of the stratification, as in nonequilibrium umbrella sampling (NEUS), accelerates convergence.  Building on ideas from MSMs and related methods, we generalize the NEUS approach in such a way that the approximation error can be reduced systematically.  We show that the improved algorithm can decrease simulation times required to achieve a desired precision by orders of magnitude.
\end{abstract}

\maketitle

\section{Introduction}

The sampling of a stochastic process can be controlled by evolving an ensemble and splitting and merging the trajectories of its members, and various algorithms based on this strategy have been introduced \cite{huber_weighted-ensemble_1996,allen_forward_2009,warmflash_umbrella_2007,dickson2010enhanced,guttenberg2012steered,bello-rivas_exact_2015,zuckerman2017weighted}.  Because the trajectory segments between splitting events are unbiased, such algorithms can yield dynamical statistics, such as transition probabilities and mean first passage times (MFPTs) to selected states, in addition to steady-state probability distributions and averages.  Furthermore, splitting algorithms generally require little communication between the members of the ensemble, making them relatively straightforward to implement regardless of the underlying dynamics, and community software is available \cite{russo_westpa_2022,lotz_wepy_2020}.  As a result, splitting algorithms are widely used.

Recent mathematical analysis of one of the oldest splitting algorithms in the molecular simulation literature, weighted ensemble (WE) \cite{huber_weighted-ensemble_1996}, shows that it is asymptotically unbiased and can dramatically reduce the variance of statistics when the splitting criteria, which are based on a partition of the state space, are chosen appropriately \cite{aristoff_weighted_2023}.  However, the method relies on convergence of the steady-state ensemble of trajectories, which is known to be slow.  In fact, as we argue below and was observed previously \cite{aristoff_weighted_2023}, it is as slow as running direct unbiased simulations.  Prior to convergence, the method is systematically
biased.

Some previous work to accelerate the convergence of WE focused on improving the initialization of walkers using an approximation of the stationary distribution.  In one case, the approximation of the stationary distribution was obtained from a free-energy method that biased trajectories to escape metastable wells \cite{ahn_gaussian-accelerated_2021}, while in another it was obtained from a machine-learning method based on unbiased short trajectories \cite{ojha_deepwest_2023}.  The former approach requires the underlying dynamics to obey detailed balance to ensure the validity of the free energy method, while the latter approach relies on a Markov approximation (but does not make other restrictions on the dynamics).  In both cases, however, the approximation to the stationary distribution is used only for the first iteration, and so if the approximation is deficient, significant bias can persist in the WE calculation for many iterations.

Others focused on improving the rate estimates from WE \cite{copperman_accelerated_2020,degrave_red_2021}.  Of particular relevance to the present study, Copperman and Zuckerman exploited the fact that splitting algorithms sample unbiased trajectory segments to construct history-augmented Markov state models (haMSMs) \cite{suarez2014simultaneous} from their WE data \cite{copperman_accelerated_2020}.  In an haMSM, the system is divided into discrete states and the probabilities for transitions between them are computed from data, just as in a standard MSM, except that the path ensemble is split based on the last metastable state visited; the rate constant is then computed as the flux from $A$ to $B$ conditioned on having last visited $A$, normalized by the steady-state probability of having last visited $A$.  The idea is that finer discretization of the state space than the WE simulation yields improved estimates.   In itself, this approach does not accelerate the convergence of the sampling.  However, Refs.\ \onlinecite{copperman_accelerated_2020,russo_westpa_2022} suggest that iteratively restarting the WE simulation from the steady-state distribution estimated from the haMSM could accelerate convergence.

 Here, we build on this idea to introduce a general framework for how to apply a short-trajectory based approximation to the stationary distribution between successive iterations.  As we show below,  the approximation must satisfy certain properties and be correctly integrated in the algorithm; if these criteria are met, the approach can accelerate the convergence by several orders of magnitude without introducing systematic bias. 
Our approach is able to compute the full stationary distribution with minimal assumptions about the dynamics, as well as kinetic statistics such as rates and committors (probabilities of visiting selected states before others).  

Conceptually, one can view our approach as a form of stratification, in which the weights of not just individual trajectory segments but groups of them are manipulated.  Stratification was first introduced for controlling the sampling of trajectory segments in nonequilibrium umbrella sampling (NEUS) \cite{warmflash_umbrella_2007,dickson2009nonequilibrium,dickson2009separating,dickson2010enhanced,dickson2011flow,dinner_trajectory_2018}. NEUS groups trajectories that are in the same regions of state space, as defined by collective variables (CVs), and estimates the steady-state probabilities of the regions by solving a global flux balance equation; this strategy was subsequently adopted in  extensions of WE \cite{bhatt_steady-state_2010} and exact milestoning (EM)\cite{bello-rivas_exact_2015}.  A unified framework for trajectory stratification that incorporates elements of all of the above algorithms and is unbiased in the limit that each region contains an infinite number of ensemble members is presented in Ref.\ \onlinecite{dinner_trajectory_2018}, which also shows that the regions can be defined in terms of path-based quantities.
Mathematical analyses of
trajectory stratification algorithms can be found in Refs.\ \onlinecite{earle2024convergence,earle2023IAD}.

Our paper is organized as follows.  In Section \ref{sec:WE_Intro}, we review the WE algorithm and recast it to show why an initialization bias is slow to disappear.  This formulation also clarifies the relation of WE to the NEUS algorithm as described in Ref.\ \onlinecite{dinner_trajectory_2018}, and we show how software for WE can be easily extended to enable trajectory stratification in Section \ref{sec:neus}.  In Section \ref{sec:badneus}, we introduce a generalization of MSMs that represents the dynamics through a basis expansion \cite{thiede_galerkin_2019, strahan_long-time-scale_2021, finkel_learning_2021} and in turn our strategy for accelerating sampling, which we term Basis-Accelerated NEUS (BAD-NEUS).   We show that the NEUS algorithm is a special case of our new scheme. In Sections \ref{sec:numex} and \ref{sec:ntl9}, we demonstrate BAD-NEUS on a two-dimensional model with a known steady-state distribution and a molecular example.

\section{Weighted Ensemble (WE)}\label{sec:WE_Intro}

As described above, in WE, the state space is partitioned, and then trajectories are copied (cloned or split) or removed (killed or pruned) from the ensemble based on criteria.  
Mathematically, we denote the state of ensemble member (henceforth, walker) $i$ at time $t$ by $X^i_t$, and we associate with $i$ a weight $w^i_t$ such that $\sum_iw^i_0=1$.
We track the index of the region containing $X^i_t$ by an index process $J^i_t$. For example, $J^i_t$ might track the element of a partition of state space in which $X^i_t$ currently resides.   However, much more general choices, such as ones based on path quantities \cite{dinner_trajectory_2018,vani_computing_2022} or distances between walkers \cite{dickson_wexplore_2014,donyapour_revo_2019}, are also possible.  See Ref.\ \onlinecite{dinner_trajectory_2018} for further discussion.

Here, we use a relatively simple procedure and represent the splitting and pruning by resampling the ensemble within each region (or, more generally, value of the index process).
Each iteration of the sampling thus consists of two steps:  evolution according to the underlying dynamics and resampling the ensemble.  In the evolution step, for each walker $i$, we numerically integrate for a time interval $\Delta$ to obtain $X^i_{t+\Delta}$ from $X^i_t$ and update $J^i_{t+\Delta}$ accordingly. In this step the weights are unchanged; $w^i_{t+\Delta}=w^i_{t}$.  We note that $\Delta$ can be a random stopping time, not just a fixed time interval.  An often useful choice is to take $t+\Delta$ to be the first time after time $t$ that the value of the index process $J^i$ changes \cite{warmflash_umbrella_2007,dickson2009nonequilibrium,dickson2009separating,dickson2010enhanced,dickson2011flow,dinner_trajectory_2018}. In this case, each walker has a different value for $\Delta.$  In the resampling step, for each value $k$ of the index process (e.g., for each region in a partition of state space), if there is at least one walker with $J^i_{t+\Delta}=k$, we select a number $N_k$. Then we sample from the set $\Phi=\{i:J^i_{t+\Delta}=k\}$ $N_k$ times with replacement according to the probabilities $p_i=w^i_{t+\Delta}/\sum_{i\in \Phi}w^i_{t+\Delta}$. For each index $r$ so chosen, we append $(X^r_{t+\Delta},J^r_{t+\Delta},\sum_{i\in \Phi}w^i_{t+\Delta}/N_k)$ to the new ensemble.

If the underlying dynamics are ergodic, in the case where $\Delta$ is a fixed time horizon, WE can be used to compute steady-state averages of functions as
\begin{widetext}
\begin{linenomath*}\begin{equation}
\mathbbm{E}_{X_0\sim \pi}[g(X_0,X_1,\ldots,X_{\Delta-1})] =\lim_{T \to \infty}\frac{1}{T}\sum_{n=0}^T\frac{1}{N}\sum_{i=0}^N g(X^i_{n\Delta},X^i_{n\Delta+1},\ldots,
X^i_{n\Delta+\Delta-1})w^i_{n\Delta},
\end{equation}\end{linenomath*}
\end{widetext}
where the subscript on the expectation indicates that we draw the initial state $X_0$ from the steady-state distribution $\pi,$ and $n$ indexes successive weighted ensemble iterations.  We present necessary modifications for the case where $\Delta$ is a random variable later.  Here and below, $X$ without a superscript indicates a realization of the underlying Markov process rather than a walker in an ensemble.
An important example is the steady-state flux into a set $D$ of interest, which can be used to compute the MFPT to $D$ from another set in the right algorithmic setup \cite{aristoff_weighted_2023}.  This flux can be obtained by setting $g(X_0,\ldots,X_\Delta)=\mathbbm{1}_{D^c}(X_0)\mathbbm{1}_D(X_1)$, where $\mathbbm{1}_D(x)$ is an indicator function that is 1 if $x\in D$ and 0 otherwise, and $D^c$ is the complement of the set $D$.  This amounts to counting the number of walkers that enter $D$ in a single time step and summing the total weight of those walkers.

Having stated the basic WE algorithm, we now present WE in a new way.  While this may initially appear to complicate the description, it reveals that WE is slow to converge and facilitates the introduction of approaches to accelerate convergence.  In this description, which we call the distribution representation of WE, we directly evolve distributions rather than approximating them through individual walkers.  To this end, we define the flux distributions
\begin{linenomath*}\begin{equation}    \overline{\pi}_\ell(dx|k)=\mathbbm{P}[X_{S(\ell)}\in dx|J_{S(\ell)}=k]
\end{equation}\end{linenomath*}
where $dx$ is an infinitesimal volume in state space that is located at a specific value of $x$, and $S(\ell)$ is an increasing sequence of stopping times. For example, if we choose $S(\ell)=\ell \Delta$ for a fixed $\Delta$, the following description corresponds to the standard version of WE already sketched. 
Alternatively, we can let $S(\ell)$ be the time of the $\ell$-th change in the value of the index process $J$, as in NEUS\cite{warmflash_umbrella_2007,dickson2009nonequilibrium,dickson2009separating,dickson2010enhanced,dickson2011flow,dinner_trajectory_2018},
\begin{linenomath*}\begin{align} \label{eq:sdef}
    S(0)&=0\\ \nonumber
    S(1)&=\min\{t>0:J_t\neq J_{t-1}\}\\\nonumber
    &\vdots\\\nonumber
    S(\ell)&=\min\{t>S(\ell-1):J_t\neq J_{t-1}\}.
\end{align}\end{linenomath*}
 This choice of stopping time has the advantage that walkers cannot go significantly beyond their regions, providing more control over the sampling.

In all practical implementations of the sampling algorithms that we discuss, the conditional flux distributions take the form of empirical distributions of $N$ walkers:  
\begin{linenomath*}\begin{equation}\label{eq:pibarestimate}
\overline{\pi}_\ell(dx|k)\approx\frac{1}{\overline{z}_\ell^k}\sum_{i=1}^Nw_{S^i(\ell)}^i\delta_{X_{S^i(\ell)}^i}(dx)\mathbbm{1}_{\{k\}}\left(J_{S^i(\ell)}^i\right),
\end{equation}\end{linenomath*}
where $\delta_x$ is the Dirac delta function centered at position $x$, and the normalization constant $\overline{z}_\ell^k$ is the total weight in region $k$ at time $S(\ell)$:
\begin{linenomath*}\begin{align}
    \overline{z}_\ell^k&=\mathbbm{P}[J_{S(\ell)}=k]
    \approx\sum_{i=1}^Nw_{S^i(\ell)}^i\mathbbm{1}_{\{k\}}\left(J_{S^i(\ell)}^i\right).
\end{align}\end{linenomath*}
The outputs of the weighted ensemble iteration and the accelerated variants that we consider here are approximations of the steady-state conditional flux distributions, as well as estimates of the corresponding normalization constants  (region weights).  
The conditional flux distributions are related to joint flux distributions through the normalization constants:
\begin{linenomath*}\begin{align}    \overline{\pi}_\ell(dx,k)&=\mathbbm{P}[X_{S(\ell)}\in dx,J_{S(\ell)}=k]\\
&= \overline{z}_\ell^k \overline{\pi}_{\ell}(dx|k),
\end{align}\end{linenomath*}
and we work with $\overline{\pi}_\ell(dx,k)$ below.

In particular, we now write the evolution and resampling steps in terms of operators.  For the former, we define the operator $\mathcal{U}$, which propagates the joint flux distribution of $(X_{S(\ell)},J_{S(\ell)})$ for a duration of $S(\ell+1)-S(\ell)$:
\begin{linenomath*}\begin{equation}
\mathcal{U}\overline{\pi}_\ell(dx,k) =  \overline{\pi}_{\ell+1}(dx,k).
\end{equation}\end{linenomath*}
In the long-time limit, this yields the eigenequation
\begin{linenomath*}\begin{equation}\label{eq:NEUS_Eig}
    \overline{\pi}(dx,k)=\mathcal{U}\overline{\pi}(dx,k),
\end{equation}\end{linenomath*}
where $\overline{\pi}$ is the steady-state joint flux distribution.
In practice, one approximates the distributions through walkers, and we denote the corresponding evolution by  
$\tilde{\mathcal{U}}$.  Mathematically,
\begin{linenomath*}\begin{multline}\label{eq:approxu}
\tilde{\mathcal{U}}\left[\frac{1}{N}\sum_{i=1}^Nw_{S^i(\ell)}^i\delta_{X_{S^i(\ell)}^i}(dx)\mathbbm{1}_{\{k\}}\left(J_{S^i(\ell)}^i\right)\right] \\
    =\frac{1}{N}\sum_{i=1}^Nw_{S^i(\ell)}^i\delta_{X_{S^i(\ell+1)}^i}(dx) \mathbbm{1}_{\{k\}}\left(J_{S^i(\ell+1)}^i\right).
\end{multline}\end{linenomath*}
The subscript on the weight does not change because the evolution step only affects the state and index of a walker, not its weight.    Operationally, this equation corresponds to propagating an ensemble's empirical distribution by running the dynamics until the stopping criterion, and then assigning the initial weight of each walker to its final state and index.
$\tilde{\mathcal{U}}$ as defined in \eqref{eq:approxu} is an unbiased stochastic approximation of $\mathcal{U}$ in the sense that, for any function $g$ and any distribution $\rho$, 
\begin{linenomath*}\begin{multline}
    \sum_k\int g(x,k)[\mathcal{U}\rho](dx,k)=\\ \mathbbm{E}\left[\sum_k\int g(x,k) [\tilde{\mathcal{U}}\rho](dx,k)\right].
\end{multline}\end{linenomath*}
If distributions are represented with some ansatz such as a neural network, one can also apply an approximate propagator based on a variational method as outlined in Refs.\ \onlinecite{wen_batch_2020,strahan_inexact_2023}.  Such a scheme would not require new trajectories to be run at each iteration, nor would it require a resampling step.

As noted above, we represent resampling through an operator, $\mathcal{R}_k$.  We define it such that, for any distribution $\rho$ and any function $g$ 
\begin{linenomath*}\begin{multline}\label{eq:Resamp_op}
    \int g(x,k) \left(\frac{1}{\zeta_k}\rho(dx,k)\right)=\mathbbm{E}\left[\int g(x,k) \mathcal{R}_k\rho\right]
\end{multline}\end{linenomath*}
and
\begin{linenomath*}\begin{equation}
    \zeta_k=\int \rho(dx,j).
\end{equation}\end{linenomath*}
Above, the LHS of \eqref{eq:Resamp_op} selects the portion of the distribution $\rho$ in region $k$ and then renormalizes it, while the RHS corresponds to resampling from the distribution.   This equation expresses the condition that the resampling operator must preserve the weighted distribution in the sense that the empirical mean of any function is the same in expectation after resampling.
A simple choice which is commonly employed for the operator $\mathcal{R}_k$ is to sample the set of walkers in region $k$ at the end of the evolution (denoted $K$) from a multinomial distribution with probability proportional to $\{w_{S^i(\ell)}^i \mathbbm{1}_{\{k\}}(J_{S^i(\ell)}^i)\}_{i\in K}$ with $N_k$ trials and then return the distribution
\begin{linenomath*}\begin{multline}\label{eq:resampop}
    \mathcal{R}_k\left[\frac{1}{N}\sum_{i=1}^N w_{S^i(\ell)}^i\delta_{X_{S^i(\ell)}^i}(dx)\mathbbm{1}_{\{k\}}(J_{S^i(\ell)}^i)\right] \\
    =\frac{1}{N_k}\sum_{i\in K}\delta_{X_{S^i(\ell)}^i}(dx)
\end{multline}\end{linenomath*}
and normalization
\begin{linenomath*}\begin{equation}
\overline{z}_\ell^k\approx\sum_i \mathbbm{1}_{\{k\}}(J_{S^i(\ell)}^i) w_{S^i(\ell)}^i.
\end{equation}\end{linenomath*}
In words, \eqref{eq:resampop} shows that ${\cal R}_k$ constructs the state distribution from a sum over the resampled walkers in each region.
There are other possible ways of resampling in the walker representation \cite{douc_comparison_2005,dickson_wexplore_2014,donyapour_revo_2019}, and these may in practice be better. However, any choice that satisfies \eqref{eq:Resamp_op} suffices for our discussion. 

Finally, we connect the flux distributions $\overline{\pi}(dx|k)$ to the steady-state distribution $\pi(x,k)$ 
for the process $(X_t,J_t)$, where the absence of superscripts again indicates the underlying Markov process rather than a walker in an ensemble.
We do this with a key identity, which we now state.  For a function $g$ of a length $\tau$ trajectory,
\begin{widetext}
\begin{linenomath*}\begin{equation}\label{eq:NEUS_Identity}
\mathbbm{E}_{(X_0,J_0) \sim \pi}[g(X_0,J_0,\dots, X_{\tau},J_{\tau})]=\frac{\sum_k\overline{z}^k \int \overline{\pi}(dx|k)\mathbbm{E}_{X_0=x,J_0=k}\left[\sum_{t=0}^{S(1)-1}g(X_t,J_t,\dots, X_{t+\tau},J_{t+\tau})\right]}{\sum_k \overline{z}^k \int \overline{\pi}(dx|k)\mathbbm{E}_{X_0=x,J_0=k}\left[S(1)\right]},
\end{equation}\end{linenomath*} 
\end{widetext}
where now $\pi$ is the steady-state joint distribution of $(X_t,J_t)$.
This identity says that, for each region $k$, we draw initial walker states $X_0$ from $\overline{\pi}(dx|k)$ and set $J_0=k$; we evolve those walkers until they leave region $k$; then we compute averages over them and weight the contribution from the walkers that started in region $k$ by $\overline{z}^k$, which is the steady-state limit of $\overline{z}_\ell^k$.
We derive \eqref{eq:NEUS_Identity} in Appendix \ref{sec:identityappendix}.
The distribution representation of WE is summarized in Algorithm \ref{alg:we}.

\begin{algorithm}[H]
\caption{WE (distribution representation)}  \label{alg:we}
\begin{algorithmic}[1]
\Require{
Approximate Markov propagator $\tilde{\mathcal{U}}$, resampling operator $\mathcal{R}_k$, starting flux distributions $\{\overline{\pi}_0(dx|k)\}_{k=1}^n$, initial region weights $\{\overline{z}_0^k\}_{k=1}^n$ with $\sum_k \overline{z}_0^k=1$, total number of iterations $L$, number of strata $n$.}
\For {$\ell = 0,\dots, L$}
    \State $\overline{\pi}_{\ell}(dx,k) \gets \overline{z}_\ell^k \overline{\pi}_{\ell}(dx|k)$
    \State $\tilde{\overline{\pi}} \gets [\tilde{\mathcal{U}} \overline{\pi}_{\ell}](dx,k)$
    \Comment{One Power Iteration Step}
    \For{$k=1, \dots, n$}
        \State $\overline{\pi}_{\ell+1}(dx|k),\overline{z}_{\ell+1}^k\gets \mathcal{R}_{k}[\tilde{\overline{\pi}}]$
    \EndFor
\EndFor
\State \Return  $\{\overline{\pi}_{L}(dx|k)\}_{k=1}^n$ and $\{\overline{z}_L^k\}_{k=1}^n$
\end{algorithmic}
\end{algorithm}
We see that each step of WE applies the operator $\mathcal{U}$ to the previous distribution $\overline{\pi}_{\ell}(dx,k)$.  Therefore, WE can be seen as performing a power iteration in ${\cal U}$, starting from some initial distribution.  The resampling step in WE plays a role analogous to the normalization step in standard power iteration in the following sense.  In standard power iteration, the normalization step serves to prevent the iterate from becoming too small or too large and introducing numerical instability.  In WE, the resampling step serves to prevent any of the region distributions from becoming poorly sampled, which would increase variance.  However, the normalization step in power iteration and, in turn, the resampling step in WE does not accelerate decay of initialization bias, nor is the WE autocorrelation time reduced relative to direct sampling.  

Algorithm \ref{alg:we} suggests that, if  $\tilde{\overline{\pi}}=\overline{\pi}$ in step 3, convergence would be achieved.  Therefore, our strategy for accelerating WE is to replace the power iteration step with a more accurate approximation.  Specifically, we compute changes of measure  (i.e., reweighting factors) $\nu_{\ell}$ such that
\begin{linenomath*}\begin{equation}\label{eq:idea}
    \overline{\pi}(dx,k) \approx \nu_{\ell}(x,k) \overline{\pi}_{\ell}(dx,k).
\end{equation}\end{linenomath*}
If we set up our approximation such that whenever $\pi_{\ell}=\pi$, we get $\nu_{\ell}=1$, then the resulting modification of WE will have a fixed point at the true steady-state distribution, and there will be no approximation error resulting from deficiencies in the approximation algorithm's specific ansatz. Our scheme has the advantage that it can potentially approach steady state  much faster than WE.  This is the idea that we develop in Sections \ref{sec:neus} and \ref{sec:badneus}.

\section{Nonequilibrium umbrellla sampling (NEUS)}\label{sec:neus}

We now present NEUS as a simple extension to WE that accelerates convergence in the way suggested in \eqref{eq:idea}; EM \cite{bello-rivas_exact_2015} can be formulated similarly.  Our development is based on the algorithm in Ref.\ \onlinecite{dinner_trajectory_2018}, which corrects a small systematic bias in earlier NEUS papers \cite{warmflash_umbrella_2007,dickson2009nonequilibrium,dickson2009separating,dickson2010enhanced,dickson2011flow}.  
We begin by making the observation that, at steady state, 
\begin{linenomath*}\begin{equation}\label{eq:NEUS_Steady_State_Condition}
\mathbbm{E}_{X_0\sim\overline{\pi}}[\mathbbm{1}_{\{k\}}(J_0)]=\mathbbm{E}_{X_0 \sim \overline{\pi}}[\mathbbm{1}_{\{k\}}(J_{S(1)})],
\end{equation}\end{linenomath*}
or, in terms of the density we wish to compute,
\begin{linenomath*}\begin{multline}\label{eq:ssobs}
    \int  \overline{\pi}(dx,k)=\\\sum_j \int \overline{\pi}(dx,j)\mathbbm{E}_{X_0=x,J_0=j}[\mathbbm{1}_{\{k\}}(J_{S(1)})].
\end{multline}\end{linenomath*}
This equation says that the steady-state probability of index process $k$ is the same as the total probability of walkers that were initialized from the steady-state distribution and flowed into (or remained in) $k$ at the stopping time.  This observation introduces one equation per possible value of the index $k$.  Therefore, we can parameterize the change of measure with as many free parameters:
\begin{linenomath*}\begin{equation}
    \overline{\pi}(dx,k) \approx  c_\ell^k \overline{z}_\ell^k \overline{\pi}_{\ell}(dx|k)
\end{equation}\end{linenomath*}
Substituting this ansatz into \eqref{eq:ssobs} and simplifying yields the matrix equation
\begin{linenomath*}\begin{equation}\label{eq:NEUS_Linear}
    c_\ell^k \overline{z}_\ell^k=\sum_jc_\ell^j \overline{z}_\ell^j\overline{G}_{jk},
\end{equation}\end{linenomath*}
where
\begin{linenomath*}\begin{equation}
    \overline{G}_{jk}=\int \overline{\pi}(dx|j)\mathbbm{E}_{X_0=x,J_0=j}[\mathbbm{1}_{\{k\}}(J_{S(1)})]
\end{equation}\end{linenomath*}
 tracks the total flux from region $j$ to region $k$.
We summarize NEUS in Algorithm \ref{alg:NEUS}.

\begin{algorithm}[H]
\caption{NEUS}
\label{alg:NEUS}
\begin{algorithmic}[1]
\Require{
Approximate Markov propagator $\tilde{\mathcal{U}}$, Resampling operator $\mathcal{R}_k$, starting flux distributions $\{\overline{\pi}_0(dx|k)\}_{k=1}^n$, initial region weights $\{\overline{z}_0^k\}_{k=1}^n$ with $\sum_k \overline{z}_0^k=1$, total number of iterations $L$, number of strata $n$.}
\For {$\ell = 0,\dots, L$}
    \State $\overline{G}_{jk} \gets \int \overline{\pi}_\ell(dx|j)\mathbbm{E}_{X_0=x,J_0=j}[\mathbbm{1}_{\{k\}}(J_{S(1)})]$
    \State Solve $c_\ell\overline{z}_{\ell}\overline{G}=c_\ell\overline{z}_{\ell}$ for $c_\ell\overline{z}_\ell$
    \State $\overline{\pi}_{\ell}(dx,k)\gets c_\ell^k\mathbbm{1}_{\{k\}}\overline{z}_\ell^k \overline{\pi}_{\ell}(dx|k)$
    \State $\tilde{\overline{\pi}} \gets \tilde{\mathcal{U}} \overline{\pi}_{\ell}(dx,k)$
    \For{$k=1, \dots, n$}
        \State $\overline{\pi}_{\ell+1}(dx|k),\overline{z}_{\ell+1}^k\gets \mathcal{R}_{k}[\tilde{\overline{\pi}}]$
    \EndFor
\EndFor
\State \Return  $\{\overline{\pi}_{L}(dx|k)\}_{k=1}^n$ and $\{\overline{z}_L^k\}_{k=1}^n$
\end{algorithmic}
\end{algorithm}

\vspace*{0.15in}

We thus see that NEUS is distinguished from WE  in its original formulation by steps 2 to 4,  in which the total fluxes between pairs of regions are estimated, and the region weights are adjusted to satisfy a global balance condition.  In practice, walkers are drawn and their dynamics are simulated to determine $S(\ell+1)$ in the loop over $\ell$ prior to step 2 (the resulting state distribution is used later in step 5); then, $\overline{G}_{jk}$ is computed in step 2 from walkers at iteration $\ell$ as
\begin{linenomath*}\begin{equation}\label{eq:NEUS_G_Data}
    \overline{G}_{jk}\approx\frac{\sum_i\mathbbm{1}_{\{j\}}\left(J^i_{S^i(\ell)}\right)\mathbbm{1}_{\{k\}}\left(J^i_{S^i(\ell+1)}\right)}{\sum_i\mathbbm{1}_{\{j\}}\left(J^i_{S^i(\ell)}\right)}.
\end{equation}\end{linenomath*}
We solve for the product $c_\ell\overline{z}_\ell$ directly and update the approximation for $\overline{\pi}$ accordingly.  We then proceed as in WE. 

 If the flux distributions and region weights assume their steady-state values such that $\overline{\pi}_\ell(dx|k)=\overline{\pi}(dx|k)$ and $\overline{z}_\ell^k=\overline{z}^k$, a global balance condition is satisfied. That is,
\begin{linenomath*}\begin{equation}\label{eq:globalbalance}
   \overline{z}^k=\sum_j \overline{z}^j\overline{G}_{jk}.
\end{equation}\end{linenomath*}
In this limit, \eqref{eq:NEUS_Linear} is solved by $c^k=1$, and NEUS has the correct fixed point (i.e., the same fixed point as $\cal U$, which encodes the unbiased dynamics).

\section{BAD-NEUS} \label{sec:badneus}

To improve on NEUS, it is necessary to improve the approximation of the change of measure in \eqref{eq:idea}.  Here we do so by introducing a basis expansion.  This allows us to vary the expressivity of the approximation through the number of basis functions, which can exceed the number of regions, in contrast to the steady-state condition in \eqref{eq:NEUS_Steady_State_Condition}.

To this end, we note that, for any lag time $\tau$ and any function $f(x,k)$, 
\begin{linenomath*}\begin{equation}
    \mathbbm{E}_{(X_0,J_0) \sim \pi}[f(X_0,J_0)-f(X_\tau,J_\tau)]=0.
\end{equation}\end{linenomath*}
We stress that this relation holds for any fixed time $\tau$.  Expanding this expectation using \eqref{eq:NEUS_Identity}, with $g(X_0,J_0,\dots,X_{\tau},J_{\tau})=f(X_0,J_0)-f(X_\tau,J_\tau)$ and multiplying through by the normalization in the denominator, we obtain
\begin{widetext}
\begin{linenomath*}\begin{equation}\label{eq:tauidentity}
0=\sum_k \overline{z}^k \int \overline{\pi}(dx|k)\mathbbm{E}_{X_0=x,J_0=k}\left[\sum_{t=0}^{S(1)-1}(f(X_t,J_t)-f(X_{t+\tau},J_{t+\tau}))\right].
\end{equation}\end{linenomath*}
We then introduce the basis set $\{\phi_{p}(x,k)\}_p$  for the change of measure in \eqref{eq:idea}, and write 
\begin{linenomath*}\begin{equation}\overline{\pi}(dx,k)=\overline{z}^k\overline{\pi}(dx|k)\approx\sum_{p} c_\ell^p\phi_{p}(x,k)\overline{z}_\ell^k\overline{\pi}_\ell(dx|k).
\end{equation}\end{linenomath*}
Making the choice $f=\phi_{r}$ and inserting the basis expansion into \eqref{eq:tauidentity} gives
\begin{linenomath*}\begin{equation}\label{eq:BAD-NEUS_1}
    0=\sum_k\sum_{p} c_\ell^p\overline{z}_\ell^k\int \phi_{p}(x,k) \overline{\pi}_\ell(dx|k) \mathbbm{E}_{X_0=x,J_0=k}\left[\sum_{t=0}^{S(1)-1}(\phi_{r}(X_t,J_t)-\phi_{r}(X_{t+\tau},J_{t+\tau}))\right].
\end{equation}\end{linenomath*}
This is a linear system which can be solved for the expansion coefficients: 
\begin{linenomath*}\begin{equation}\label{eq:BADNEUS_Solve}
0=\sum_{p}c_\ell^pM_{pr}.
\end{equation}\end{linenomath*}
with the matrix entries given by
\begin{linenomath*}\begin{equation}\label{eq:BAD-NEUS_M}
    M_{pr}=\sum_k\overline{z}_\ell^k\int \phi_{p}(x,k) \overline{\pi}_\ell(dx|k) \mathbbm{E}_{X_0=x,J_0=k}\left[\sum_{t=0}^{S(1)-1}(\phi_{r}(X_t,J_t)-\phi_{r}(X_{t+\tau},J_{t+\tau}))\right].
\end{equation}\end{linenomath*}
\end{widetext}
The basis set must include the constant function in its span so that the system has a unique nontrivial solution.

\begin{algorithm}[H]
\caption{BAD-NEUS}\label{alg:badneus}
\begin{algorithmic}[1]
\Require{
Approximate Markov propagator $\tilde{\mathcal{U}}$, Resampling operator $\mathcal{R}_k$, starting flux distributions $\{\overline{\pi}_0(dx|k)\}_{k=1}^n$, initial region weights $\{\overline{z}_0^k\}_{k=1}^n$ with $\sum_k \overline{z}_0^k=1$, basis set $\{\phi_{p}(x,k)\}_p$, lag time $\tau$, number of past iterations to retain $h$, total number of iterations $L$, number of strata $n$.}
\For {$\ell = 0,\dots, L$}
    \State $\overline{\pi}_\ell \gets \overline{\pi}_{\ell,h}$
    \State $\Delta\phi\gets\sum_{t=0}^{S(1)-1}\phi_{r}(X_t,J_t)-\phi_{r}(X_{t+\tau},J_{t+\tau})$
    \State $M_{pr} \gets \sum_{k}\int \overline{z}_\ell^k\phi_{p}(x,k) \overline{\pi}_{\ell}(dx|k)\mathbbm{E}_{X_0=x,J_0=k}\left[\Delta\phi\right]$
    \State Solve $cM=0$
    \State $\overline{\pi}_{\ell}(dx,k)\gets \sum c_\ell^p \phi_p(x,k)\overline{z}_\ell^k\overline{\pi}_{\ell}(dx|k)$
    \State $\tilde{\overline{\pi}} \gets \tilde{\mathcal{U}} \overline{\pi}_{\ell}(dx,j)$
    \For{$k=1, \dots, n$}
        \State $\overline{\pi}_{\ell+1}(dx|k),\overline{z}_{\ell+1}^k\gets \mathcal{R}_{k}[\tilde{\overline{\pi}}]$
    \EndFor
\EndFor
\State \Return  $\{\overline{\pi}_{L}(dx|k)\}_{k=1}^n$ and $\{\overline{z}_L^k\}_{k=1}^n$
\end{algorithmic}
\end{algorithm}

We summarize the BAD-NEUS algorithm in Algorithm \ref{alg:badneus}.
If the flux distributions and region weights assume their steady-state values such that $\overline{\pi}_\ell(dx|k)=\overline{\pi}(dx|k)$ and $\overline{z}_\ell^k=\overline{z}^k$, then 
\eqref{eq:tauidentity} is satisfied for all $f$.  In that case, just as \eqref{eq:globalbalance} and \eqref{eq:NEUS_Linear} lead to  $c^k=1$ in NEUS, \eqref{eq:tauidentity} and \eqref{eq:BAD-NEUS_1} lead to $\sum_p c_p \phi(x,k)=1$ in BAD-NEUS. In this case, step 6 in Algorithm \ref{alg:badneus} reduces to $\overline{\pi}_{\ell+1}(dx,k)\gets \overline{z}^k\overline{\pi}(dx|k)$, and BAD-NEUS has the correct fixed point.

 Specific choices for the basis functions, lag time, and stopping time correspond to existing methods.
NEUS as described in Section \ref{sec:neus} and Algorithm \ref{alg:NEUS} corresponds to the choice
\begin{linenomath*}\begin{equation}
    \phi_{p}(x,k)=\mathbbm{1}_{\{p\}}(k),
\end{equation}\end{linenomath*}
$\tau=1$,  and $\Delta$ as defined in \eqref{eq:sdef}.   We can obtain a version of iteratively restarted WE as suggested in Refs.\ \onlinecite{copperman_accelerated_2020,russo_westpa_2022} by choosing the basis set to be indicator functions that are one on sets that are finer than the stratification (``microbins''), $\tau=1$, and a fixed stopping time $\Delta$. For these choices, the sum over $t$ in \eqref{eq:BAD-NEUS_M} telescopes, and $M$ can be row-normalized and written as $T-I$, where $T$ is the transition matrix for the MSM defined by the basis set.

Consistent with WE and NEUS, in practice, we estimate the integral $M_{pr}$ through a sum over walkers:
\begin{widetext}
\begin{linenomath*}\begin{equation}\label{eq:BAD_NEUS_M_Data}
M_{pr}\approx\sum_k\sum_{i}\overline{z}_\ell^k\phi_{p}(X^i_{S^i(\ell)},k)\mathbbm{1}_{\{k\}}(J^i_0)\left[\sum_{t=S^i(\ell)}^{S^i(\ell+1)-1}(\phi_{r}(X^i_{t},J_{t}^i)-\phi_{r}(X^i_{t+\tau},J_{t+\tau}^i))\right].
\end{equation}\end{linenomath*}
\end{widetext}
Given $M$ and, in turn, the estimated expansion coefficients, we update the weights as
\begin{linenomath*}\begin{equation}\label{eq:weightFormula}
    w_{S^i(\ell)}^i=\frac{\overline{z}_\ell^{J_{S^i(\ell)}^i}}{Z}\sum_{p} c_\ell^p \phi_{p}\left(X_{S^i(\ell)}^i,J_{S^i(\ell)}^i\right),
\end{equation}\end{linenomath*}
where $Z$ normalizes the total of the walker weights to one.    Operationally, we loop over each walker before the resampling step and evaluate \eqref{eq:weightFormula} to adjust its weight.  Physically, this corresponds to apportioning the weight of the region containing the walker ($\overline{z}_\ell^{J_{S^i(\ell)}^i}/Z$ in \eqref{eq:weightFormula}) according to the basis expansion representing the change of measure to the steady-state.  We then proceed with the resampling step as in the usual WE algorithm.

\begin{figure*}
\begin{minipage}{\linewidth}
\begin{algorithm}[H]
\caption{Trajectory Stratification (walker representation)}  
\label{alg:sampling_walkers}
\begin{algorithmic}[1]
\Require{
Starting ensemble of walkers $E_0=\{(X_0^i,J_0^i,w_0^i)\}_{i=1}^N,$ with $\sum_{i=1}^N w_0^i=1$,  lag time $\tau$, number of past iterations to retain $h$, total number of iterations $L$, number of strata $n$.
}
\For {$\ell = 0,\dots, L$}
    \State $N \gets \text{length}(E_\ell)$
    \State Generate a list of trajectories $T_{\ell} \gets \{[(X_{S^i(\ell)}^i,J_{S^i(\ell)}^i,w_{S^i(\ell)}^i),(X_{S^i(\ell)+1}^i,J_{S^i(\ell)+1}^i,w_{S^i(\ell)}^i),\dots,$\par \hfill$(X_{S^i(\ell+1)+\tau}^i,J_{S^i(\ell+1)+\tau}^i,w_{S^i(\ell)}^i)]\}_{i=1}^N$,  where $(X_{S^i(\ell)}^i,J_{S^i(\ell)}^i,w_{S^i(\ell)}^i)=E_{\ell}[i]$
    \State $T_{\ell} \gets \text{concatenate}(\{T_{\ell},T_{\ell-1},\dots T_{\ell-h+1}\})$
    \State $N \gets \text{length}(T_\ell)$
    \State Renormalize the weights in $T_\ell$ by dividing each by $h$.
    \State Update weights $\{w^i_{S^i(\ell)}\}_{i=1}^N$ using Algorithm \ref{alg:Approx}. (Omit for Weighted Ensemble)
    \State Initialize an empty list $E_{\ell+1}=\{\}$
    \For{$k = 1, \dots, n$}
        \State $\overline{z}^k_{\ell+1} \gets \sum_{i=1}^N w_{S^i(\ell)}^i\mathbbm{1}_{\{k\}}(J^i_{S^i(\ell+1)})$
        \For{$r=1, \dots, N_k$}
            \State Sample an index $b$ with probability proportional to $w_{S^b(\ell)}^b \mathbbm{1}_{\{k\}}(J^b_{S^b(\ell+1)})$
            \State Append to $E_\ell$ the configuration $(X_{S^b(\ell+1)}^b,J_{S^b(\ell+1)}^b,\overline{z}^k_{\ell+1}/N_k)$
        \EndFor
    \EndFor
\EndFor
\State \Return  $T_L$ and $\{\overline{z}_L^k\}_{k=1}^n$
\end{algorithmic}
\end{algorithm}
\end{minipage}
\end{figure*}

\begin{figure*}
\begin{minipage}{\linewidth}
\begin{algorithm}[H]
\caption{Approximating the steady-state flux distribution (walker representation)}  \label{alg:Approx}
\label{alg:flux_walkers}
\begin{algorithmic}[1]
\Require{
List of trajectories $T_\ell$, lag time $\tau$, basis functions $\{\phi_p\}_p$, region weights $\{\overline{z}_\ell^k\}_{k=1}^n$
}
\State $M_{pr} \gets \sum_{ki}\overline{z}_\ell^k\phi_{p}(X^i_{S^i(\ell)},k)\mathbbm{1}_{\{k\}}(J^i_0)\left[\sum_{t=S^i(\ell)}^{S^i(\ell+1)-1}(\phi_{r}(X^i_{t},J_{t}^i)-\phi_{r}(X^i_{t+\tau},J_{t+\tau}^i))\right]$

\State Solve $c_{\ell}M=0$ 

\State $w_{S^i(\ell)}^i \gets (\overline{z}_\ell^{J_{S^i(\ell)}^i}/Z)\sum_{p} c_\ell^p \phi_{p}\left(X_{S^i(\ell)}^i,J_{S^i(\ell)}^i\right)$
\State \Return $\{ w_{S^i(\ell)}^i \}_{i=1}^N$
\end{algorithmic}
\end{algorithm}
\end{minipage}
\end{figure*}

To reduce variance in our estimate of the matrix $M$ to ensure a stable solve, it is often useful to work with the mixture distribution for the last $h$ iterations:
\begin{linenomath*}\begin{equation}
    \overline{\pi}_{\ell,h}(dx|k)=\frac{1}{h}\sum_{t=\ell-h+1}^\ell\overline{\pi}_t(dx|k).
\end{equation}\end{linenomath*}
When working with walkers, this mixture distribution corresponds to concatenating the walkers from the last $h$ iterations, and therefore contains more data than using a single iteration.  We simply substitute $\overline{\pi}_{\ell,h}$ for $\overline{\pi}_\ell$ in all of our algorithms.  A unified walker-based algorithm that we use in practice is given in Algorithm \ref{alg:sampling_walkers}, with the additional steps needed for NEUS and BAD-NEUS in Algorithm \ref{alg:Approx}.

Finally, we note that the general strategy of improving the approximation of the change of measure is not limited to using a basis expansion.  Just as one can use various means to solve the equations of the operator that encodes the statistics of the dynamics \cite{thiede_galerkin_2019,strahan_long-time-scale_2021,strahan_predicting_2023,strahan_inexact_2023}, one can represent the change of measure here by either a basis expansion or a nonlinear representation.  In particular, our tests based on the neural-network approach in Ref.\ \onlinecite{strahan_inexact_2023} show significant promise (unpublished results).

\section{Two-dimensional test system}\label{sec:numex}

\begin{figure}[bt]
\begin{center}
\includegraphics[scale=.45]{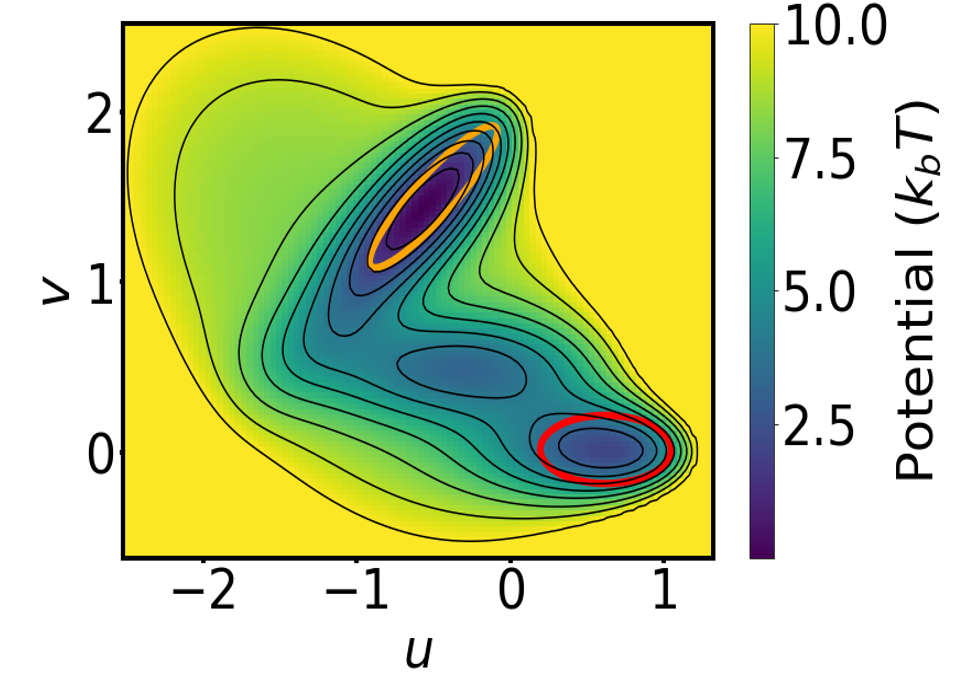}
\end{center}
\caption{\label{fig:MB}
M\"uller-Brown potential.  Orange and red ovals indicate states $A$ and $B$, respectively.  Contours are spaced every $k_BT$
}
\end{figure}

We first illustrate our approach by sampling a two-dimensional system, which enables us to compare estimates of the steady-state distribution to the Boltzmann probability to ensure that simulations are run until convergence.  Specifically, the system $x=(u,v)$ is defined by the M\"uller-Brown potential \cite{muller_location_1979}, which is a sum of four Gaussian functions:
\begin{linenomath*}\begin{multline}\label{eq:MB}
V_{\rm MB}(u,v)=\frac{1}{20}\sum_{i=1}^4C_i \exp[a_i(u-v_i)^2\\+b_i(u-u_i)(v-v_i)+c_i(v-v_i)^2].
\end{multline}\end{linenomath*}
For all results shown, we use $C_i=\{-200, -100, -170, 15\}$, $a_i=\{-1,-1,-6.5,0. 7\}$,
$b_i=\{0,0,11,0.6\}$, 
$c_i=\{-10,-10,-6.5,0. 7\}$, 
$u_i=\{1,-0.27,-0.5,-1\}$, 
$v_i=\{0,0.5,1.5,1\}$.
The potential with these parameter choices is shown in Figure \ref{fig:MB}.  The presence of multiple metastable states and a minimum energy pathway that does not parallel the axes of the variables used for the numerical integration make this system representative of difficulties commonly encountered in molecular simulations. Because the model is two-dimensional, we can readily visualize results and compare them with statistics independently computed using the grid-based scheme in Ref.\ \onlinecite{strahan_inexact_2023}.  

We evolve the system with overdamped Langevin dynamics, discretized with the BAOAB algorithm \cite{leimkuhler_rational_2013}: 
\begin{linenomath*}\begin{equation}\label{eqn:MBEuler}
X_{t+dt}=X_t-\nabla V_{\rm MB}(X_t)dt+\sqrt{\frac{dt}{2\beta}}(Z_t+Z_{t-dt}),
\end{equation}\end{linenomath*}
where $dt$ is the time step, $\beta$ is the inverse temperature, and $Z_t\sim N(0,I)$ is a random vector with components drawn from the normal distribution with zero mean and unit standard deviation ($I$ is the two-dimensional identity matrix). For all results shown, we use  $dt=0.001$ and $\beta=2$.

\subsection{Comparison of algorithms}

A key point of our theoretical development is that NEUS and BAD-NEUS are simple elaborations of WE.  This enables us to define a unified walker-based algorithm that we use in practice (Algorithm \ref{alg:sampling_walkers}, with the added operations needed for NEUS and BAD-NEUS in Algorithm \ref{alg:Approx}).
Splitting and stratification are defined through the rule for switching the index process.  The sequence of stopping times $S(\ell)$ is then determined by the index process through \eqref{eq:sdef}. The update rule for $J_t$ that we employ for all three algorithms is as follows.  Let $\{\psi_k(x)\}_{k=1,\ldots, n}$ be a set of nonnegative functions. If $\psi_{J_t}(X_{t+dt})>0$, then $J_{t+dt}=J_t$; otherwise, $\mathbbm{P}[J_{t+dt}=k]=\psi_k(X_{t+dt})/\sum_k\psi_k(X_{t+dt})$.  That is, the value of the index process remains the same until the walker leaves the support of $\psi_{J_t}$, and then a new index value $k$ is drawn with likelihood proportional to $\psi_k(X_{t+dt})$.  For the M\"uller-Brown model, we use
\begin{linenomath*}\begin{equation}
\psi_k(u,v) = \begin{cases}
1& \text{if}\ |v-v_k|<\varepsilon_k\ \text{and}\ 1<k<n\\ 
1& \text{if}\ v-v_k<\varepsilon_k\ \text{and}\ k=1\\
1& \text{if}\ v_k-v>\varepsilon_k\ \text{and}\ k=n\\
0& \text{otherwise}.
\end{cases}
\end{equation}\end{linenomath*}
Unless otherwise indicated, we set $n=10$, space the $v_k$ uniformly in the interval $[-0.2,1.8]$, and set $\varepsilon_k=0.6(v_{k+1}-v_{k})$, so that the regions of support (strata) overlap.  
This choice prevents walkers in barrier regions from rapidly switching back and forth between index values, limiting the overhead of the algorithms. 
Unless otherwise indicated, we use 2000 walkers per region and run them until their index processes switch values.  Initial configurations for $J_0=k$ are generated by uniformly sampling the support of $\psi_k$.  

For BAD-NEUS, we use a basis set consisting of 10 indicator functions per stratum. The indicator functions are determined by clustering all the samples in a stratum with $k$-means clustering and partitioning it into Voronoi polyhedra based on the cluster means.  
Unless otherwise indicated, we compute statistics using data from the last $h=3$ iterations and use a lag time of $\tau=10$ time steps.  

\begin{figure}[bt]
\begin{center}
\includegraphics[scale=.45]{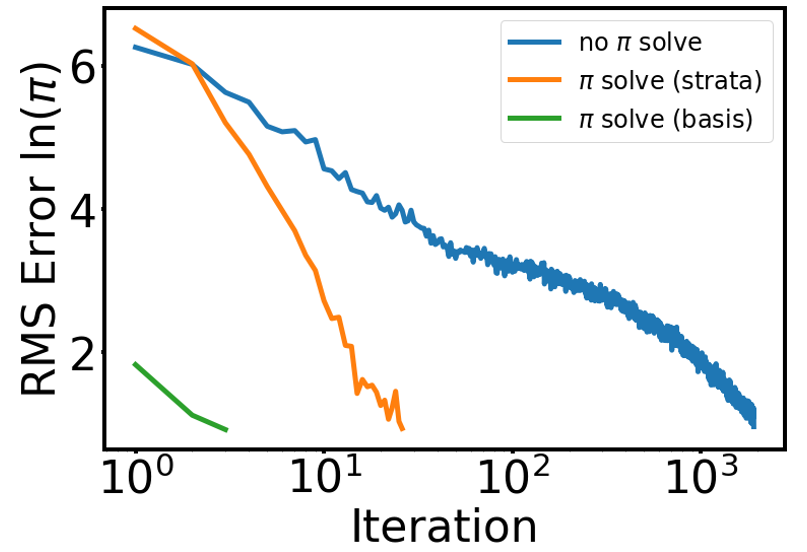}
\end{center}
\caption{
Convergence of the steady-state distribution of the M\"uller-Brown system for WE as defined in Algorithm 1 (blue, no $\pi$ solve), NEUS as defined in Algorithm 2 (orange, $\pi$ solve using the strata in (37) as a basis), and BAD-NEUS as defined in Algorithm 3 (green, $\pi$ solve using a basis defined by $k$-means clustering, as described in the text).  
}
\label{fig:MB_NEUS}
\end{figure}

We measure convergence by computing the root mean square (RMS) difference in $\ln(\tilde{\pi}(x))$ from $-\beta V_{\rm MB}(x)$, where $\tilde{\pi}(x)$ is an estimate of the steady-state distribution from the normalized histogram of samples.  For the histogram, we use a uniform $50\times 50$ grid on the rectangle defined by the minimum and maximum values in the NEUS dataset.  Since some grid regions are empty because they correspond to energies too high for NEUS to sample, we only compute errors over bins with $V_{\rm MB}<7$.  We stop each simulation when the RMS difference drops below one.  Results are shown in Figure \ref{fig:MB_NEUS}.  We see that WE  (without any acceleration strategy) requires about 73 times more iterations than NEUS, which in turn requires about 13 times more iterations than BAD-NEUS to reach the convergence criterion.

\begin{figure}[bt]
\includegraphics[scale=.7]{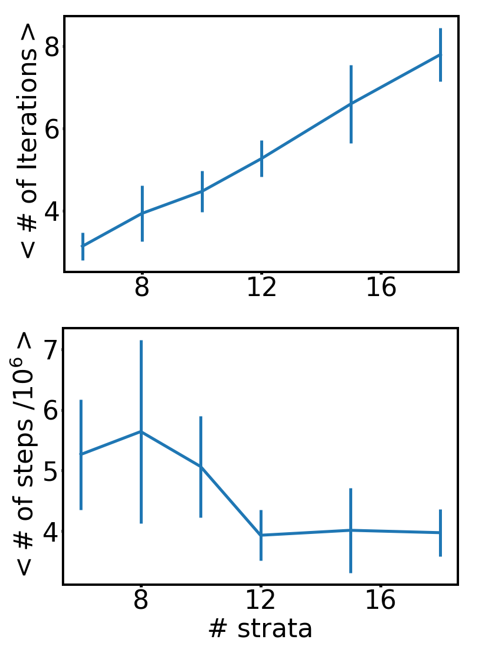}
\caption{
Effect of the number of strata on convergence of BAD-NEUS applied to the M\"uller-Brown system. The number of iterations required (top) and the total number of time steps (bottom)---a proxy for computational effort---until convergence.  The total number of walkers is held fixed.  Error bars are the standard deviation over 10 replicate simulations.
For these simulations, there are 40,000 walkers total, and the lag time is $\tau=1$ time step.
}
\label{fig:MB_Strata}
\end{figure}

\subsection{Choice of hyperparameters}

Having demonstrated that trajectory stratification outperforms WE (without any acceleration strategy) for this system, we now examine the effect of key hyperparameters in NEUS and BAD-NEUS.

First, we investigate the effect of the number of strata. To compare simulations that require essentially the same resources, we hold the total number of walkers fixed at 40,000, and we divide the walkers uniformly across the strata. For these simulations, we use a lag time of $\tau=1$ and retain $h=3$ past iterations.  In Figure \ref{fig:MB_Strata} we plot both the number of iterations to reach convergence and the total number of time steps needed for convergence as we vary the number of strata.   We see that the number of BAD-NEUS iterations required for convergence increases linearly with the number of strata. However, the lengths of trajectories decrease because the strata are spaced more closely. Due to this interplay, the total effort to achieve convergence decreases until about 12 strata and then levels off.  While actual performance will depend on the overhead incurred by stopping and starting the dynamics engine and computing the BAD-NEUS weights, these results indicate that finer stratification is better.

\begin{figure}[b]
\includegraphics[scale=.72]{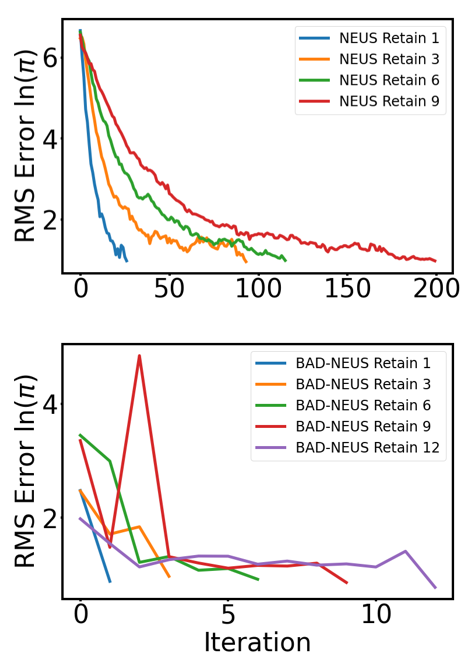}
\caption{\label{fig:retention}
Effect of retaining past iterations on convergence of NEUS (top) and BAD-NEUS (bottom) applied to the M\"uller-Brown system.
}
\end{figure}

Next, we examine the dependence of convergence on the number of iterations used to compute statistics, $h$ (other hyperparameters are set to the default values given above).
Using data from a larger number of iterations allows for more averaging, but it can also contaminate the statistics with samples obtained before the steady-state distribution has converged.
While the trends are clearer for NEUS than BAD-NEUS because the former converges more slowly, Figure \ref{fig:retention} indicates that retaining fewer iterations is better for both NEUS and BAD-NEUS.  It is therefore important to use enough walkers per iteration that data from  past iterations need not be retained.  If one has access to a large number of processing elements, and many walkers can be simulated in parallel, this is not a severe restriction.  Once convergence is achieved one can begin accumulating data to reduce variance.

\begin{figure}[b]
\includegraphics[scale=.435]{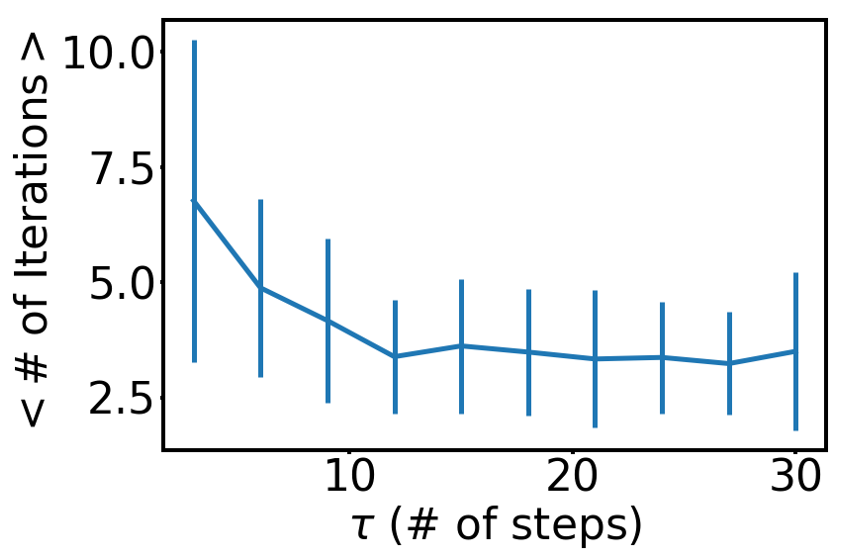}
\caption{
Dependence of convergence on the lag time for BAD-NEUS applied to the M\"uller-Brown system.  Error bars are the standard deviation over 10 replicate simulations.
}
\label{fig:Lag_BAD}
\end{figure}

Finally, we investigate the impact of the lag time in Figure \ref{fig:Lag_BAD} (other hyperparameters are set to the default values given above).  We find that using longer lag times weakly decreases the number of iterations required for convergence.

\subsection{Kinetic statistics}

A common problem in molecular dynamics is that while equilibrium averages are relatively straightforward to calculate from biased simulations (via, for example, the various methods reviewed in Ref.\ \onlinecite{henin2022enhanced}), dynamical averages are much harder.  In this section, we use BAD-NEUS to compute dynamical averages efficiently.  To do so, we split  the ensemble of transition paths based on the last metastable state visited as previously for NEUS \cite{vanden2009exact,dickson2009separating,vani_computing_2022}. We specifically compute the backward committor, $q_-$, which is the probability that the system last visited a reactant state $A$ rather than a product state $B$, and the transition path theory (TPT) rate, defined as the mean number of $A$ to $B$ transitions per unit time divided by the fraction of time spent last in $A$.
The former can be obtained without saving additional information once we split the path ensemble.  The forward committor, $q_+$, which is the probability that the system next visits state $B$ rather than state $A$, can also be obtained from NEUS but requires saving additional information \cite{vani_computing_2022}.  Here, both of the systems that we consider are microscopically reversible, so we can obtain $q_+$ from $q_+=1-q_-$.

For the M\"uller-Brown system, we consider the following history dependent stratification.  Let $\{\psi_{k}^A(x)\}_{k=1,\ldots, n}$ and $\{\psi_{k}^B(x)\}_{k=1,\ldots, m}$ be non-negative functions, let $\psi_k=\psi_{k}^A$ if $k\leq n$; otherwise, let $\psi_k=\psi_{k-n}^B,$ and let $\mathbbm{1}_A$ and $\mathbbm{1}_B$ be indicator functions on sets $A$ and $B$, respectively.  Let $T^-_{A\cup B}(t)$ be the time the system sampled at time $t$ was last in $A$ or $B$, so that $\mathbbm{1}_A(X_{T^-_{A\cup B}(t)}) = 1-\mathbbm{1}_B(X_{T^-_{A\cup B}(t)})$ reports whether $A$ rather than $B$ was last visited.  Define the index process by the following update rule:

\begin{widetext}
\begin{linenomath*}\begin{equation}\label{eqn:historysplit}
\mathbbm{P}[J_{t+dt}=k]=
\begin{cases}
\psi_{k}(X_{t+dt})\mathbbm{1}_A(X_{T^-_{A\cup B}(t+dt)})/\sum_i\psi_i^A(X_{t+dt}) & k\leq n, \psi_{J_t}(X_{t+dt})=0  \text{ or } \\ & \quad\quad\quad\mathbbm{1}_A(X_{T^-_{A\cup B}(t+dt)})\neq \mathbbm{1}_A(X_{T^-_{A\cup B}(t)})\\ 
\psi_{k}(X_{t+dt})\mathbbm{1}_B(X_{T^-_{A\cup B}(t+dt)})/\sum_i\psi_i^B(X_{t+dt}) & k > n ,\psi_{J_t}(X_{t+dt})=0  \text{ or }\\ & \quad\quad\quad \mathbbm{1}_A(X_{T^-_{A\cup B}(t+dt)})\neq \mathbbm{1}_A(X_{T^-_{A\cup B}(t)})\\ 
\mathbbm{1}_{\{k\}}(J_{t}) & \text{otherwise}
\end{cases}
\end{equation}\end{linenomath*}
\end{widetext}
    
Thus there are $n+m$ total strata, the first $n$ of which contain walkers that last visited $A$, and the remainder of which contain walkers that last visited $B$.  The steady-state distribution of walkers which last visited $A$ is proportional to $\pi(x)q_{-}(x)$, and that which last visited $B$ is proportional to $\pi(x)(1-q_-(x))$.  This ensemble lets us compute dynamical averages.  The backward committor projected onto a space of CVs can be computed using
\begin{linenomath*}\begin{equation}\label{eq:NEUS_qb}
    q_{-}^{\theta}(s)=\frac{\mathbbm{E}[\mathbbm{1}_{\{0,\ldots, n\}}(J_t)\mathbbm{1}_{\{ ds\}}(\theta(X_t))]}{\mathbbm{E}[\mathbbm{1}_{\{ds\}}(\theta(X_t))]},
\end{equation}\end{linenomath*}
where $\theta$ is a vector with components that are the CVs and $ds$ is a bin in the space.
The TPT rate constant is defined as
\begin{linenomath*}\begin{equation}\label{eqn:tstrate}
k_{AB} =\frac{\E_{\pi}[q_{-}(x)\mathbbm{1}_{B^c}(x)\mathcal{T}_1\mathbbm{1}_{B}(x)]}{\E_{\pi}[q_{-}(x)]},
\end{equation}\end{linenomath*}
 where $B^c$ is the complement of $B$, and $\mathcal{T}_t$ is the operator that describes the evolution of expectations of functions:
\begin{linenomath*}\begin{equation}\label{eqn:transitionoperator}
    \mathcal{T}_t g(x,j) = E_{X_0=x,J_0=j}[g(X_t,J_t)].
\end{equation}\end{linenomath*}
In practice, we compute \eqref{eqn:tstrate} from \eqref{eq:NEUS_Identity} by choosing
\begin{linenomath*}\begin{multline}
    g(X_0,J_0 ..., X_t,J_t)=\\\mathbbm{1}_{B^c}(X_0)\mathbbm{1}_{B}(X_1)\mathbbm{1}_{\{0,\ldots,n\}}(J_0)
\end{multline}\end{linenomath*}
for the numerator and 
\begin{linenomath*}\begin{equation}
    g(X_0,J_0 ..., X_t,J_t)=\mathbbm{1}_{\{0,\ldots,n\}}(J_0)
\end{equation}\end{linenomath*}
for the denominator.  
This corresponds to the steady-state flux into $B$ from trajectories that originated in $A$.

For the M\"uller-Brown system, we define states $A$ and $B$ as
\begin{widetext}
\begin{linenomath*}\begin{equation}
\label{eqn:MBstates}
\begin{aligned}
&A=\set{u, v : 6.5(u+0.5)^2-11(u+0.5)(v-1.5)+6.5(u-1.5)^2<0.3}\\
&B=\set{u, v:(u-0.6)^2+0.5(v-0.02)^2<0.2}.
\end{aligned}
\end{equation}\end{linenomath*}
\end{widetext}
We use a similar index process construction as for the equilibrium calculations, except there are 10 $v_k$ that are evenly spaced in the interval $[0,1.6]$ for $\psi_{k}^A$ and 10 $v_k$ that are evenly spaced in the interval $[-0.3,0.8]$ for $\psi_{k}^B$, for a total of 20 strata.  We use different definitions for $\psi_{k}^A$ and $\psi_{k}^B$ because, for walkers originating from state $A$ ($B$), a stratum located below (above) state $B$ ($A$) is unlikely be populated.  Both NEUS and BAD-NEUS use 2000 walkers per stratum, and we compute statistics using data from the last $h=3$ iterations.  For BAD-NEUS, we use a basis set consisting of 10 indicator functions per stratum, again based on $k$-means clustering; we use a lag time of $\tau=1$.  

To compute reference values for these kinetic statistics, we use the grid-based approximation to the generator in Ref.\ \onlinecite{strahan_inexact_2023}, with the same grid parameters.  Since the dynamics are microscopically reversible, we obtain the backward committor by solving for the forward committor using the approach in Ref.\ \onlinecite{strahan_inexact_2023}, then setting $q_-=1-q_+$.  We solve for the reaction rate using 
\begin{linenomath*}\begin{equation} \label{eq:gridrate}
k_{AB}=\frac{2(\vec{\pi q}_-)^\top(P\vec{\mathbbm{1}}_B)}{(\vec{\pi}^\top \vec{q}_-)(\beta \epsilon_x)},
\end{equation}\end{linenomath*}
where $P$ is the discretized transition matrix defined on the grid, $\epsilon_x$ is the grid spacing, and arrows indicate vectors of function values on the grid points ($\vec{\pi q}_-$ is a single vector of product values).  

\begin{figure}[bt]
\includegraphics[scale=.4]{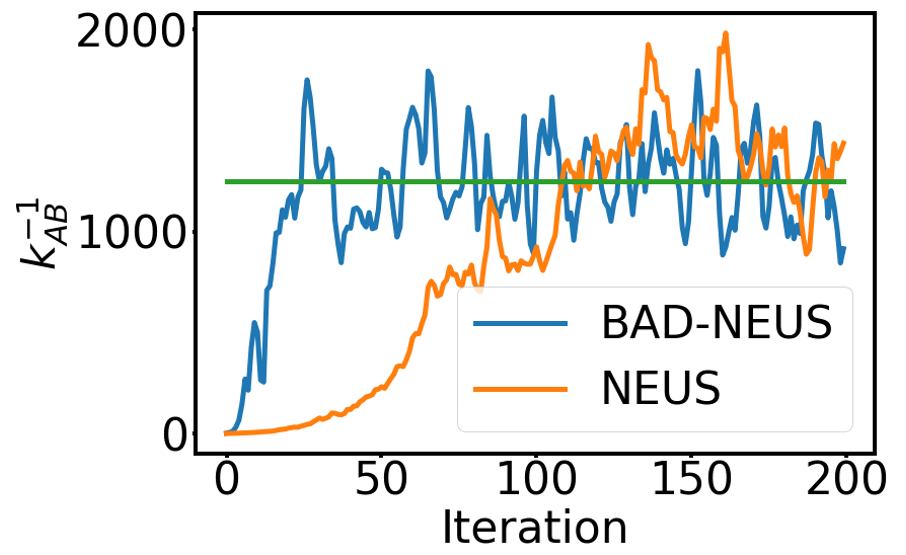}
\caption{
Convergence of the TPT inverse rate estimate for the M\"uller-Brown system.  The green line is the grid-based reference obtained from \ref{eq:gridrate}.
For these simulations, there are 2000 walkers per stratum, $h=4$ iterations are retained, and the lag time is $\tau=10$ time steps.}
\label{fig:MB_TPT_Rate}
\end{figure}

Figure \ref{fig:MB_TPT_Rate} shows estimates for the TPT rate.  
We see that BAD-NEUS converges several fold faster than NEUS, 
Each BAD-NEUS iteration requires an average of 407 time units of sampling (arising from 2000 walkers in each of 20 strata, with an average time before leaving the stratum of 10.2 time steps).  Each iteration is therefore significantly less total computational cost than generating a single $A$ to $B$ transition on average (1200 time units, as evidenced by $k_{AB}^{-1}$) and is amenable to parallelization.

\begin{figure}[bt]
\includegraphics[scale=.34]{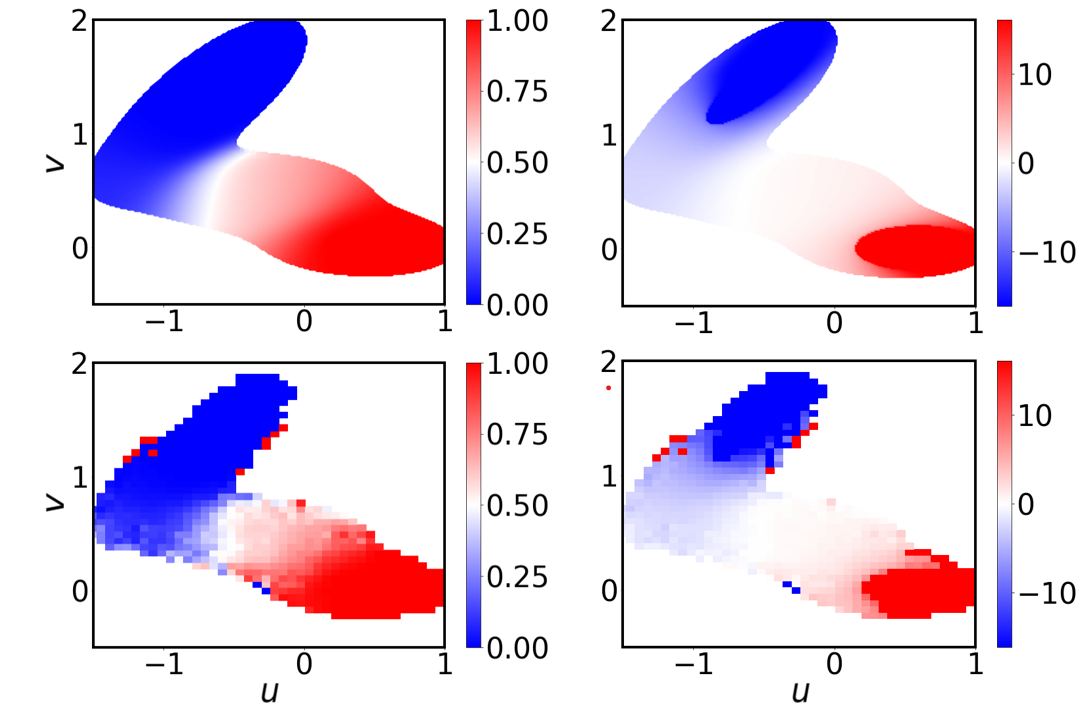}
\caption{
Illustration of the backward committor computed from BAD-NEUS.  (top) Reference.  (bottom) BAD-NEUS results. (left) Backward committor.  (right) $\ln(q_-/(1-q_-))$, which emphasizes states near $A$ and $B$. These results are obtained from the same simulations as Fig.\ \ref{fig:MB_TPT_Rate}.
}
\label{fig:MB_Q}
\end{figure}

Figure \ref{fig:MB_Q} shows the backward committor computed at BAD-NEUS iteration 20 using \eqref{eq:NEUS_qb} with the variables of numerical integration as the CVs.  We represent the results in two ways:  the committor itself and its logit function.  The former emphasizes the transition region ($q_-\approx 0.5$) while the latter emphasizes values close to states $A$ and $B$ ($q_-\approx 0$ and $q_-\approx 1$). Both show excellent agreement with the reference.

\section{Molecular Test System}\label{sec:ntl9}

\begin{figure}[bt]
\centering
\includegraphics[width=0.45\textwidth]{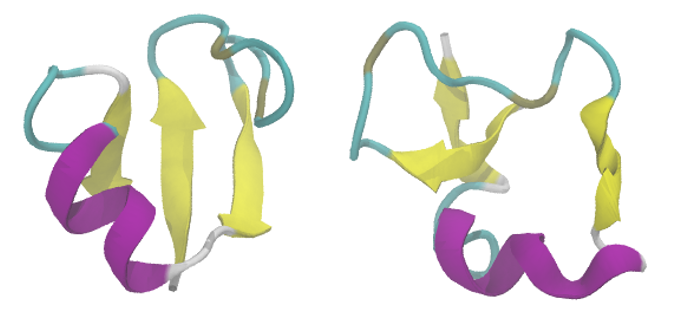}
\caption{
Representative NTL9(1-39) structures drawn from the (left) native and (right) denatured states in the BAD-NEUS simulation. 
There is residual helical structure in the denatured state in both our simulations and those to which we compare \cite{adhikari_computational_2019}.}
\label{fig:NTL9_Struct}
\end{figure}

As a more challenging test of the method, we estimate the folding rate of the first 39 amino acids of the N-terminal domain of ribosomal protein L9 (NTL9). The native secondary structure of NTL9 is $\beta\beta\alpha\beta\alpha$ sequentially.  In the native state, the $\beta$-strands form an anti-parallel $\beta$-sheet, and the $\alpha$-helices pack on either side of it.  NTL9(1-39) lacks the C-terminal helix (Fig.\ \ref{fig:NTL9_Struct}).  We choose this system because it enables direct comparison with an earlier computational study  \cite{adhikari_computational_2019} (described further below), which estimated the folding to be on the millisecond timescale, consistent with experimental studies of (full) NTL9 \cite{luisi_effects_1999,sato_n-terminal_2017} and earlier simulations of NTL9(1-39) \cite{voelz2010molecular}.

All molecular dynamics simulations were performed with OpenMM \cite{eastman_openmm_2017}.  We compute CVs including backbone RMS deviations (RMSDs) and fractions of native contacts using MDtraj \cite{McGibbon2015MDTraj}.  To enable direct comparison with Ref.\ \onlinecite{adhikari_computational_2019}, we model NTL9(1-39) using the AMBER FF14SB force field \cite{maier_ff14sb_2015} and the Hawkins, Cramer and Truhlar generalized Born implicit solvent model \cite{hawkins_parametrized_1996,hawkins_pairwise_1995}.  We use a Langevin thermostat with a time step of 2 fs, a temperature of 300. K, and a friction constant of $\gamma=80$ ps$^{-1}$, which corresponds to the high-viscosity case considered in Ref.\ \onlinecite{adhikari_computational_2019}.

We run the molecular dynamics for each walker in intervals of 20 ps and compute the CVs to update the index process at the end of each interval.  For BAD-NEUS, after an index process changes values (($S(1)$ in \eqref{eq:sdef}), we run the walker $\tau$ additional molecular dynamics steps to ensure that we have a total of $\tau$ steps beyond $S(1)$.  While this procedure allows walkers to run beyond the time that they exit their stratum, it does not bias the results.  Since our algorithm only requires us to stop and start the molecular dynamics engine while running unbiased dynamics, we do not need to modify OpenMM in any way. 

The stratification is similar to that for the M\"uller-Brown system.  Namely, we set
\begin{linenomath*}\begin{equation}
\psi_{k}^A(X) =\psi_{k}^B(X) =\begin{cases}
1& \text{if}\ |Q(X)-Q_{k}|<\varepsilon_k\\ 
0& \text{otherwise},
\end{cases}
\end{equation}\end{linenomath*}
where $Q(X)$ is the fraction of native contacts using the definition of Ref.\ \onlinecite{best_native_2013}.  We determine the native contacts from the crystal structure in the Protein Data Bank (PDB), PDB ID 2HBB \cite{cho_energetically_2014}. We space 20 $Q_{k}$ values uniformly in the interval $[0.35,0.85]$ and set $\varepsilon_k=0.6(Q_{k+1}-Q_{k})$, so that the regions overlap.  Thus there are 40 total strata.
We define the folded state as full backbone RMSD $ < 0.28$ nm and $Q> 0.85$; the denatured state has full backbone RMSD $>0.80$ nm and $Q <0.35$.  

To initialize the simulation, we minimize the energy starting from the PDB structure, draw velocities from a Maxwell-Boltzmann distribution for 380 K, and simulate for 80 ns at that temperature; this denatures the protein (Fig.\ \ref{fig:NTL9_Struct}).  We sample 4000 frames in each of the 40 strata from this trajectory and weight them uniformly; this set forms our starting walkers, $\{X^i_0,J^i_0,1/N\}_{i=1}^N$, where $N=40$.  We use a lag time of 10 ps, and we retain a history of $h=4$ iterations.  The initialization pipeline, integrator settings, and stratification choice are the same between NEUS and BAD-NEUS.    The hyperparameter choices are summarized in Table \ref{tab:ntl9params}.

\begin{table}[tb]
    \centering
    \caption{Hyperparameter choices used for NTL9(1-39) simulations}
    \begin{tabular}{lc}
     \hline
        \textbf{Hyperparameter}
         & \textbf{Kinetic statistics}
          \\ \hline 
         Stratification CV &  $Q$ and $\mathbbm{1}_A(T^-_{A\cup B}(t))$ \\
         Stopping rule & Stratum exit\\
         Walkers per stratum  & 4000\\
         Number of strata  & 40\\
         Number of iterations retained, $h$  & 4\\
         Type of basis set & $k$-means indicator\\ 
         Number of basis functions per stratum & 10\\ 
         Lag time, $\tau$  & 10 ps\\
        \hline
    \end{tabular}
    \label{tab:ntl9params}
\end{table}

To construct the basis set for BAD-NEUS, we define the following seven CVs: (1) the fraction of native contacts, (2) the full backbone RMSD, (3-5) the backbone RMSD for each of the three $\beta$-strands taken individually (residues 1-4, 17-20, and 36-38), (6) the backbone RMSD for the $\alpha$-helix  (residues 22-29), and (7) the backbone RMSD for the three $\beta$-strands together.  Our basis set consists of 10 indicator functions on each stratum constructed by $k$-means clustering on the seven-dimensional CV space.
At each iteration (i.e., one pass through the outer loop in Algorithm \ref{alg:sampling_walkers}), we determine the basis functions for stratum $j$ by taking the associated cluster centers from the prior iteration and refining them with 10 iterations of Lloyd's algorithm \cite{lloyd_least_1982} using all the samples with index process $J_t^i=j$ from the last $h$ iterations.  

Here we focus on the convergence of the folding rate; a detailed analysis of the NTL9 folding mechanism (with and without the C-terminal helix) based on potentials of mean force and committors will be presented elsewhere (see also Ref.\ \onlinecite{strahan2024short}).  As mentioned above, we compare our results to those of Ref.\ \onlinecite{adhikari_computational_2019}, which estimates rates using haMSMs applied to data from WE (haMSM-WE). As discussed in the Introduction, BAD-NEUS goes beyond haMSM-WE by using the basis set (Markov model) to accelerate the convergence of the sampling rather than just the rate estimates. 

\begin{figure}[bt]
\begin{center}
\includegraphics[scale=.35]{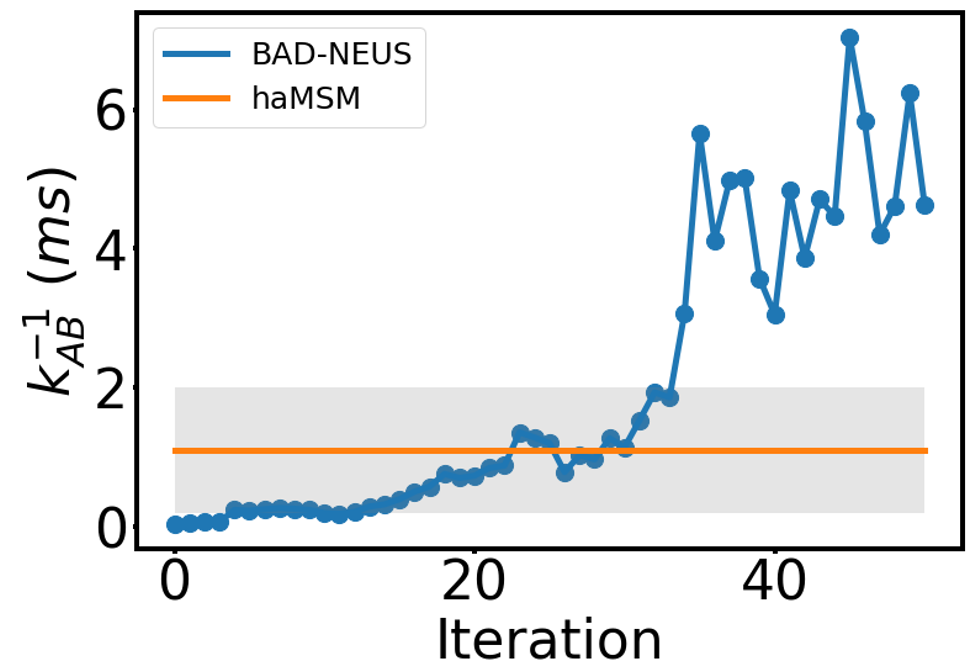}
\caption{
Comparison of inverse rates for NTL9(1-39) folding from BAD-NEUS and haMSM-WE. The haMSM-WE results are from Ref.\ \onlinecite{adhikari_computational_2019} The shaded area shows the Bayesian $95\%$ credible interval reported in that publication. 
}
\label{fig:NTL9_Rate}
\end{center}
\end{figure}

Figure \ref{fig:NTL9_Rate} shows the inverse rate constant obtained from BAD-NEUS and haMSM-WE \cite{adhikari_computational_2019}. 
The average inverse rate over the final 15 iterations of BAD NEUS is 4.74 ms, with a standard deviation of 0.99 ms, while the authors of Ref.\ \onlinecite{adhikari_computational_2019} used a Bayesian bootstrapping approach to estimate a confidence interval of 0.17--1.9 ms, which they state is likely an underestimate of the true 95 percent confidence interval owing to the limited number of independent samples used in the analysis.  Given that we have only a single BAD NEUS run, the standard deviation of the final 15 (highly correlated) inverse rate estimates is also likely a significant underestimate of our statistical error.  We thus view the results in Figure \ref{fig:NTL9_Rate} as in agreement.

\begin{figure}[bt]
\begin{center}
\includegraphics[scale=.70]{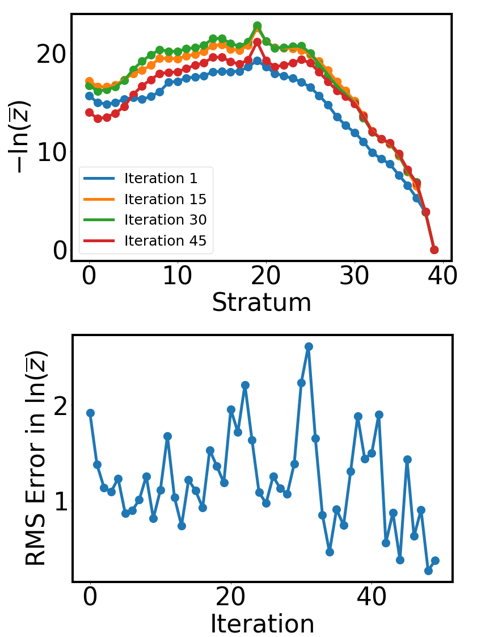}
\caption{
(top) BAD NEUS bin weights for the NTL9(1-39) system.  Strata with indices less than 20 correspond to the folding direction (trajectories last in the denatured state), and strata with indices greater than or equal to 20 correspond to the  unfolding direction (trajectories last in the native state).  (bottom) Deviation of weights ($\ln(\overline{z})$) from those in the last iteration.  
}
\label{fig:NTL9_zbars}
\end{center}
\end{figure}

We examine the weights in more detail in Figure \ref{fig:NTL9_zbars}.  The increase in the inverse rate around iteration 35 in Figure \ref{fig:NTL9_Rate} (and concommitant decrease in below 5 in Figure \ref{fig:NTL9_zbars}(bottom)) corresponds to a shift in the weights of strata with indices close to 25 (compare iterations 30 and 45 in Figure \ref{fig:NTL9_zbars}(top)).  These strata contain walkers that are nearly denatured from trajectories that last visited the native state, and the shift in weights in that region stabilizes the denatured state relative to the native state.

 We run each walker until its index process changes and then an additional 10 ps, so the trajectory length and time per iteration are random variables. For the NTL9 simulations, the average trajectory length was 50 ps, and the average time per iteration was 4 $\mu$s.  The results in Figure \ref{fig:NTL9_Rate} thus required about 204 $\mu$s in total, comparable to the 252 $\mu$s for the haMSM-WE calculation. The time to compute the rate constant by BAD-NEUS and haMSM-WE is much less than the millisecond timescale of a single folding event, which we also expect to be the time to converge WE without any acceleration strategy. BAD-NEUS actually provides some speedup relative to haMSM-WE in the sense that it provides potentials of mean force and committors in its present form, whereas a calculation in the unfolding direction (performed either separately or simultaneously, as here) would be required to obtain these statistics from haMSM-WE. By the same token, we likely could reduce the computation time for BAD-NEUS by ``teleporting'' walkers from $B$ to $A$ and computing only the forward rate \cite{baudel2023hill}.  However, we expect most investigators would seek the mechanistic information provided by the potentials of mean force and committor (which can be used to also obtain reactive currents \cite{strahan_long-time-scale_2021}) if they are investing the computational resources to compute the rate, and free energies computed from ratios of forward and backward rates can serve as checks on the rates \cite{hall2020calculating}.

\section{Conclusions}
Trajectory stratification methods enable enhanced sampling for estimating both thermodynamic (equilibrium) and kinetic (nonequilibrium) statistics.
Here, we introduced a new trajectory stratification method, BAD-NEUS, which converges faster than existing ones.  We show that our method is a natural generalization of  NEUS and EM, which in turn can be formulated as extensions of WE.  In the process, we show that WE as originally formulated \cite{huber_weighted-ensemble_1996} converges no faster than unbiased dynamics. 
The key modification introduced in NEUS and elaborated here is the insertion of an approximation to the steady-state distribution before the resampling step.   Importantly, we design this approximation algorithm so that it preserves the steady-state distribution of the dynamics, and therefore does not introduce systematic errors into the algorithm in the large-data limit.

Our approach defines precise ingredients needed for iterative restart strategies to converge to the correct fixed point (i.e., a fixed point consistent with unbiased dynamics).  In this way, it both restricts and generalizes 
iterative restart strategies previously proposed \cite{copperman_accelerated_2020,russo_westpa_2022}, 
and we provide numerical demonstrations that such strategies can accelerate convergence of both equilibrium and kinetic statistics.
In the future, it would be interesting to compare BAD-NEUS with the WE milestoning approach, which instead embeds WE simulations within
coarse regions (i.e., between milestones) \cite{ray_weighted_2020,ray_markovian_2022}.  While this approach also combines fine and coarse information, it is fundamentally different from BAD-NEUS and haMSM-WE in that it introduces an additional sampling rather than analysis algorithm at the fine level. Indeed, adding an element of stratification to the WE simulations within WE milestoning may result in further speedups.
The similarities and differences between various approaches highlights the importance of a theoretical framework for relating algorithms like the one that we introduce here; we believe it can serve as the basis for a thorough, systematic cataloging of methods for computing dynamical statistics, analogous to recent attempts for enhanced sampling methods for equilibrium statistics \cite{henin2022enhanced}.

In the present study, we use a simple basis expansion to model the steady-state distribution, but our strategy is general, and alternatives are also possible.  One could use neural networks to learn the steady-state distribution and/or the basis functions \cite{wen_batch_2020,strahan_inexact_2023}.
One could also incorporate memory \cite{lorpaiboon2024accurate}.  These alternatives, which could be used separately or together, alleviate the need to identify basis sets  and/or CVs that describe the dynamics well and, in the case of memory, may reduce the sampling required.   We thus expect our method to be a significant step toward accurate estimates of rates for protein folding and similarly complex molecular conformational changes.

\begin{acknowledgements}
We thank Daniel Zuckerman for discussions concerning the manuscript. This work was supported by National Institutes of Health award R35 GM136381 and National Science Foundation award DMS-2054306.
This work was completed in part with resources provided by the University of Chicago Research Computing Center.
``Beagle-3: A Shared GPU Cluster for Biomolecular Sciences'' is supported by the National Institutes of Health (NIH) under the High-End Instrumentation (HEI) grant program award 1S10OD028655-0.\\
\end{acknowledgements}

\appendix

\section{Derivation of \eqref{eq:NEUS_Identity}}
\label{sec:identityappendix}
Here we derive \eqref{eq:NEUS_Identity} by modifying the proof from Ref.\ \onlinecite{moustakides_extension_1999}.  Let $Z_t$ be a Markov process, $\mathcal{T}_t$ be operator defined in \eqref{eqn:transitionoperator}, 
and $S_n=\sum_{t=0}^{n-1}g(Z_t)$ be the partial sum of a function $g$.  Let $\omega(z)$ satisfy the linear problem
\begin{linenomath*}\begin{equation}
    (\mathcal{T}_1-\mathcal{I})\omega(z)=-(g(z)-\mathbbm{E}_{\pi}[g(z)])
\end{equation}\end{linenomath*}
subject to the constraint $\mathbbm{E}_{\pi}[\omega]=0$.

Consider the process:
\begin{linenomath*}\begin{equation}
    U_n=S_n-\mathbbm{E}_{\pi}[g]n +\omega(Z_n)
\end{equation}\end{linenomath*}
Then $U_t$ is a martingale with respect to $Z_0, ..., Z_t$.  To see this, we compute:
\begin{widetext}
\begin{linenomath*}\begin{align}
    \mathbbm{E}[U_{t+1}-U_{t}|Z_0,...Z_t]&=\mathbbm{E}[S_{t+1}-\E_{\pi}[g](t+1)+\omega(Z_{t+1})-\left(S_{t}-\mathbbm{E}_{\pi}[g]t+\omega(Z_{t}) \right)|Z_0,...Z_t] \nonumber\\
    &=\mathbbm{E}[g(Z_t)-\E_{\pi}[g]+\omega(Z_{t+1})-\omega(Z_{t})|Z_0,...Z_t]\nonumber\\
    &=g(Z_t)-\E_{\pi}[g]+\mathbbm{E}[\omega(Z_{t+1})-\omega(Z_{t})|Z_0,...Z_t]\nonumber\\
    &=g(Z_t)-\E_{\pi}[g]+(\mathcal{T}_1-\mathcal{I})\omega(Z_t)\nonumber\\
    &=g(Z_t)-\E_{\pi}[g]-g(Z_t)+\mathbbm{E}_{\pi}[g]\nonumber\\
    &=0.
\end{align}\end{linenomath*}
We refer the reader to Ref.\ \onlinecite{moustakides_extension_1999} for a more technical discussion and conditions under which the relevant expectations are bounded such that we may apply the optional stopping theorem.  Then, for any stopping time $T$ with $\E[T]<\infty$,
\begin{linenomath*}\begin{equation}
    0=\E[U_T-U_{0}]=\E[S_T]-\E_{\pi}[g]\E[T]+\E[\omega(Z_T)]-\E[\omega(Z_0)],
\end{equation}\end{linenomath*}
which yields the result:
\begin{linenomath*}\begin{equation}\label{eq:NEUS_Identity_Exact}
\frac{\E[\sum_{t=0}^{T-1}g(Z_t)]}{\E[T]} =\E_{\pi}[g]+\frac{\E[\omega(Z_0)]-\E[\omega(Z_T)]}{\E[T]}.
\end{equation}\end{linenomath*}
Now take $Z_t=(X_t,J_t, ..., X_{t+\tau},J_{t+\tau})$ and take the stopping time to be $S(1)$. In the case where $X_0,J_0$ is distributed according to the steady state flux distribution $\overline{\pi}(dx,k),$ \eqref{eq:NEUS_Eig} implies that $(X_T,J_T) \sim \overline{\pi}(dx,k)$, and hence the distribution of $Z_0$ and $Z_T$ are the same, and so the residual term in \eqref{eq:NEUS_Identity_Exact} is zero.  In this case, we note that the distribution of $(X_0,J_0)$ is given by:
\begin{linenomath*}\begin{align}\label{eq:LOTP}
\pi(dx)&=\mathbbm{P}[X_{0}\in dx]\\
&=\sum_k\mathbbm{P}[X_{0}\in dx\mid J_{0}=k]\mathbbm{P}[J_{0}=k]\\
&=\sum_k\overline{z}^k\overline{\pi}(dx|k).
\end{align}\end{linenomath*}
The joint distribution for $(X,J)$ is $\overline{\pi}(dx,k)=\overline{z}^k\overline{\pi}(dx|k)$.  Therefore, 

\begin{linenomath*}\begin{equation}
\mathbbm{E}\left[\sum_{t=0}^{S(1)-1}g(X_t,J_t, ..., X_{t+\tau},J_{t+\tau})\right]=
\sum_k \overline{z}^k \int \overline{\pi}(dx|k)\mathbbm{E}_{X_0=x,J_0=k}\left[\sum_{t=0}^{S(1)-1}g(X_t,J_t, ..., X_{t+\tau},J_{t+\tau})\right]
\end{equation}\end{linenomath*}
and
\begin{linenomath*}\begin{equation}
\mathbbm{E}[S(1)]=\sum_k \overline{z}^k \int \overline{\pi}(dx|k)\mathbbm{E}_{X_0=x,J_0=k}[S(1)].
\end{equation}\end{linenomath*}
\end{widetext}
Substituting these into \eqref{eq:NEUS_Identity_Exact} and remembering that the residual term is zero yields \eqref{eq:NEUS_Identity}.


\end{document}